\RequirePackage{fix-cm}
\documentclass[twocolumn,epjc3]{svjour3}  
\smartqed  
\RequirePackage{graphicx}
\RequirePackage{amsmath}
\usepackage{lineno}
\usepackage{wasysym}

\RequirePackage[colorlinks,citecolor=blue,urlcolor=blue,linkcolor=blue]{hyperref}
\journalname{Eur. Phys. J. A}
\begin{document}


\title{Production and characterization of a PbMoO$_4$ cryogenic detector from archaeological Pb}

\subtitle{}
\author{    L.~Pattavina\thanksref{LNGS,GSSI,e1} \and
  S.~Nagorny\thanksref{LNGS,e2} \and
       S.~Nisi\thanksref{LNGS} \and
       L.~Pagnanini\thanksref{LNGS} \and
       G.~Pessina\thanksref{MIB} \and
      S.~Pirro\thanksref{LNGS} \and
      C.~Rusconi\thanksref{LNGS,USC} \and
      K.~Sch\"affner\thanksref{LNGS,GSSI}
      V.N. Shlegel\thanksref{Novos,Novos1}
      V.N.~Zhdankov\thanksref{fabbrica}
}

\thankstext{e1}{Corresponding authors: luca.pattavina@lngs.infn.it and sn65@queensu.ca}
\thankstext{e2}{Present address: Queen's University, K7L 3N6 Kingston, Canada}

\institute{INFN - Laboratori Nazionali del Gran Sasso, Assergi I-67100 - Italy\label{LNGS}
\and
Gran Sasso Science Institute, L'Aquila I-67100 - Italy\label{GSSI} 
\and
INFN - Sezione di Milano-Bicocca, I-20126 Milano - Italy\label{MIB}
\and
Department of Physics and Astronomy, University of South Carolina, SC-29208 Columbia - USA\label{USC}
\and
Nikolaev Institute of Inorganic Chemistry, 630090 Novosibirsk - Russia\label{Novos}
\and
Novosibirsk State Tech University, 20 Karl Marx Prospect, 630092 Novosibirsk - Russia\label{Novos1}
 \and
CML Ltd., 630090 Novosibirsk - Russia\label{fabbrica}
}

\date{Received: date / Accepted: date}

\maketitle

\begin{abstract}
We operated a PbMoO$_4$ scintillating cryogenic detector of 570~g, produced with archaeological lead. This compound features excellent low temperature characteristics in terms of light yield, 12~keV/MeV for $\beta/\gamma$ interactions, and FWHM energy resolution, 11.7~keV at 2.6~MeV. Furthermore, the detector allows for an effective particle identification by means of pulse shape analysis on the heat read-out channel. The implementation of innovative techniques and procedures for the purification of raw materials used for the crystal growth, and the highly-pure archaeological Pb, allowed the production of large volume high quality crystal. The overall characteristics of the detector operated at cryogenic temperatures makes PbMoO$_4$ an excellent compound for neutrino physics applications, especially neutrinoless double-beta studies.
\end{abstract}

\section{Introduction}
In the last few decades increasing consideration has been given to the study of neutrinoless double-beta decay (0$\nu\beta\beta$):
\begin{equation}
(A,Z) \rightarrow (A,Z+2)+2e^-.
\end{equation}
This rare weak process has never been observed until now, but its survey has strong and direct implications in many fields of physics, both in the neutrino sector and beyond the standard model. The observation of 0$\nu\beta\beta$ decay can prove the existence of massive Majorana neutrinos and, if the signal was strong enough to measure the half-life of the transition, the absolute mass scale of neutrinos together with their mass hierarchy could be constrained~\cite{Ponte}. The process could also shed light on the asymmetry between matter and antimatter in the Universe~\cite{Lepto}.

Neutrinoless double-beta decay can occur only in the nuclei with an even number of both protons and neutrons. One of the most interesting candidate nuclei is $^{100}$Mo. This nucleus is characterized by a large Q$_{\beta\beta}$-value (3035~keV~\cite{100Mo}) and it is available in nature in different compounds, like: CaMoO$_4$, ZnMoO$_4$ and Li$_2$MoO$_4$. For these reasons, this isotope is of great interest for bolometric ultra-low background investigations~\cite{IHE,LMO_exp,AMORE}.

However, among the Mo-based crystal a promising but not yet fully exploited one is PbMoO$_4$. This compound has one of the lowest concentrations in Mo, only 26\% of its mass is made of Mo (in the case of Li$_2$MoO$_4$ is 55\%). Nevertheless, thanks to its high density the concentration of Mo moles is 0.019~mol/cm$^3$ (0.017~mol/cm$^3$ for Li$_2$MoO$_4$). 

Using a PbMoO$_4$ crystal as a thermal detector may unleash a great discovery potential for neutrino physics, namely 0$\nu\beta\beta$ decay. In fact, using the low temperature cryogenic technique, an optimal detection efficiency (at the level of 90\%)~\cite{MC_bolometer}, excellent energy resolution in the region of interest (ROI)~\cite{ZnMoO4_big} and outstanding discrimination of the interacting particle~\cite{CMO} are ensured.

Furthermore, Pb is a very promising target element for neutrino physics in general, especially for the investigation of Coherent Elastic Neutrino-Nucleus Scattering~\cite{7even}, given its favourable cross section. In fact, in this process the interaction cross section scales as the number of neutrons in target squared. The nuclear stability of Pb is also exploited for the detection of Supernova neutrinos. The HALO~\cite{SN} experiment uses Pb as target material for the detection of neutrino coming from astrophysical sources.

In this work we characterized a PbMoO$_4$ cryogenic detector, made from archaeological Pb, for 0$\nu\beta\beta$ investigations.

\section{Mo-based compounds}
Among the different compounds with Mo, the ones which were widely tested and demonstrated their suitability for bolometric double-beta investigations are: ZnMoO$_4$~\cite{ZnMoO4_big}, Li$_2$MoO$_4$~\cite{LiMoNe,LMO_exp} and CaMoO$_4$~\cite{CMO}. Nevertheless, they still need to prove their reproducibility, in terms of crystal quality and radiopurity, for a future ton-scale cryogenic detector (using about 1000 channels), as the one proposed by CUPID~\cite{CUPID_1,CUPID_2}. Lowering the background for the investigation of 0$\nu\beta\beta$ is of paramount relevance, given the elusive counting rate of the decay source. The two main requirements that must be met by future experiments are: the availability of high-quality enriched crystals, so to increase the number of $\beta\beta$ nuclei under investigation and thus to enhance the sensitivity, and the high internal radiopurity of the crystals as discussed in~\cite{IHE}.


In this paper we present the performance of a scintillating calorimetric cryogenic detector of PbMoO$_4$. This crystal is of great potential. A drawback is the almost unavoidable presence of the radioactive isotope $^{210}$Pb, within modern lead, which would give a high intrinsic background. In fact, if the raw starting materials for the crystal production are thoroughly selected, the crystal radiopurity may meet the needed requirements. Nevertheless reducing the $^{210}$Pb in natural Pb metal samples is extremely arduous. In this context, only the use of archaeological lead may help in getting around this problem. The PbMoO$_4$ crystal studied in this work, was, in fact grown using archaeological lead~\cite{4mBq}. This is not the first time a compound produced from archaeological Pb is studied. In 2013, our group operated for the first time a PbWO$_4$ scintillating bolometer for the investigation of Pb isotopes $\alpha$ decay~\cite{pbwo4}. Also, in 2017 we tested the cryogenic performance of a small mass PbMoO$_4$ crystal~\cite{YRM}.

\section{Production of archaeological PbMoO$_4$ crystals}
In the following we describe the two most relevant processes needed for the production of high quality single PbMoO$_4$ crystals: synthesis of the raw materials and crystal growth.

\subsection{Raw materials for crystal production}
In order to produce a high quality crystal generally reagents with the highest available purity should be used. Nevertheless, it is often necessary to make a compromise between crystal quality and cost of the crystal production. However, special care should always be taken to minimize the content of transition metals (Cr, Mn, Fe, Co, Ni). The concentration of each those elements should not exceed 10$^{-6}$~g/g, because higher concentrations would significantly degrade the optical quality, reduce the light output and worsen the energy resolution as well as the bolometric properties of the crystal.

On the other side, archaeological Pb is not a material available on the market. For this reason a small piece of Pb ingot from a sunken ship of the Roman empire ages, near Sardinia~\cite{Alessandrello}, was used for our studies. A purification process is needed prior to the use of archaeological Pb for crystal production, due to the low initial purity, 99.9\%. In our case, the vacuum distillation method, which was developed and described in detail in~\cite{vacuum} was used for the purification. After the vacuum distillation the concentration of the major impurities in the raw Pb metal (Sb, Sn and Ag) are reduced more than 10-500 times (see Tab.~\ref{Tab1}). The mass fraction of Pb after refining is 99.9996\%, that fully complies with the chemical purity requirements of reagents for high quality crystal production.

In the next step, lead oxide (PbO) powder was produced from the refined Pb metal by means of several sequential chemical transformations: dissolution of purified Pb in weak nitric acid solution followed by neutralization by the acid solution by gaseous ammonia, and lead hydroxide Pb(OH)$_2$ precipitation. This substance was washed several times with ultra-pure water and centrifuged. The beta modification of lead oxide ($\beta$-PbO), which is used for synthesis of PbMoO$_4$ charge, was done annealing the lead hydroxide in a quartz crucible gradually increasing the temperature up to 820$^\circ$K, for a total duration of about 24~h. The resulting PbO powder met the requirements of chemical purity for the production of high quality PbMoO$_4$ crystals. The chemical purity of the material used for the crystal production for the different processing steps is shown in Tab.~\ref{Tab1}.

The production of PbMoO$_4$ powder was carried out directly into a Pt crucible of 70~mm diameter and 150~mm height dedicated for this crystal growth. In order to prevent the imbalance of PbMoO$_4$ charge components due to MoO$_3$ evaporation during the crystal growth, a small excess of MoO$_3$ (1.0~\%) was added to the stoichiometric mixture of PbO and MoO$_3$ powders. The modification of the stoichiometry balance caused by the different vapour pressure of  MoO$_3$ and PbO (in the temperature range from 773~K to 1273~K) is the main issue for the production of high quality PbMoO$_4$ single crystals~\cite{Ruski}. The mixture of charge components was heated up to the melting point (1370~K), with a temperature gradient of 100~K/h. The total duration of the homogenisation and synthesis processes was about 10 hours.
	
The contamination level of the archaeological lead before and after the refining, the lead oxide, the synthesized PbMoO$_4$ charge, and the grown PbMoO$_4$ crystal was monitored using a High Resolution Inductively Coupled Plasma-Mass Spectrometer (HR-ICP-MS, Thermo Fisher Scientific ELEMENT2) at the Laboratori Nazionali del Gran Sasso of INFN (LNGS, Italy). The results of the analyses are listed in Tab.~\ref{Tab1}. After the crystal growth the element with the largest contamination in was W. The increase in the W concentration in the crystal is due to a contamination of the Pt crucible used for previous growth. In fact, as described later in the text, the same crucible was also used in the past for the production of other crystal containing W (e.g. ZnWO$_4$, PbWO$_4$).

\begin{table*}[t]
\centering
\begin{tabular}{|c|c|c|c|c|c|}
\hline
Element & Pb before        & Pb after         & PbO powder       & MoO$_3$ powder    & PbMoO$_4$ crystal    \\ \hline
Cr      & 200    & {$<$}40    & {$<$}5000    &  {$<$}2000  &{$<$}5000     \\ \hline
Mn      & {$<$}30    & {$<$}50    & {$<$}500    & 200            & {$<$}500    \\ \hline
Fe      & 200 & {$<$}30 & {$<$}30 & {$<$}2400           & {$<$}250000  \\ \hline
Co      & {$<$}30    & {$<$}30     & {$<$}250     & {$<$}500           & {$<$}250     \\ \hline
Ni      & {$<$}40   & {$<$}40    & {$<$}2500    & {$<$}30000          & {$<$}2500    \\ \hline
Cu      & {$<$}20   &{$<$}20      & {$<$}3000   &  {$<$}10000 & {$<$}3000              \\ \hline
Cd      & {$<$}20    & {$<$}300    & {$<$}1500    & {$<$}100  &  {$<$}1500     \\ \hline
Zn      & {$<$}300   & {$<$}70      & {$<$}7000   & 7500              & {$<$}7000   \\ \hline
As      & {$<$}70            & {$<$}250   & {$<$}1000   & {$<$}1000 & {$<$}1000   \\ \hline
Ag      & 160000            &{$<$}200        & {$<$}2500    & {$<$}1000 & {$<$}2500    \\ \hline
Sn      & 1000000           & 5200   & {$<$}2500   & {$<$}1000 & {$<$}2500    \\ \hline
Sb      & 250000            & 2600    & {$<$}13000    & {$<$}1000& {$<$}13000    \\ \hline
W       & {$<$}700   & {$<$}700   & {$<$}2500   & 47000   & 160000            \\ \hline
Th      & {$<$}300     & {$<$}300     & {$<$}100     & {$<$}50    & {$<$}100     \\ \hline
U       & {$<$}300     & {$<$}300     & {$<$}100     & {$<$}50   &      {$<$}100      \\ \hline
\end{tabular}
\caption{Impurity concentration in Pb samples: before and after vacuum distillation, of the produced PbO, synthetized PbMoO$_4$ powder and of the final crystal. The values are measured with High Resolution Inductively Coupled Plasma-Mass Spectrometer (HR-ICP-MS, Thermo Fisher Scientific ELEMENT2) at the Laboratori Nazionali del Gran Sasso of INFN (LNGS, Italy), and the units are in 10$^{-9}$~g/g.}
\label{Tab1}
\end{table*}

\subsection{Crystal growth}

The conventional Czochralski growth was used to produce the first archaeological PbMoO$_4$ single crystal described in~\cite{YRM}. The low-background measurement of a 57~g PbMoO$_4$ crystal, operated as scintillating cryogenic detector demonstrated the high level of contamination of $^{238}$U/$^{232}$Th decay chain products at the level of Bq/kg. This internal contamination was probably due to re-contamination during crystal growth process.
This time, to prevent any possible re-contamination the Low Thermal Gradient (LTG) Czochralski technique~\cite{LTG1,LTG2,LTG3} was used, which operates with a closed Pt-crucible. An additional and very important advantage of this technique is the small temperature gradient used, which is at the level of about 1~K/cm. Thanks to this innovative approach, there is no intensive evaporation of the charge components from the melt, and losses of the initial materials are extremely small, less than 0.5~\%. Last, but not least, the LTG Czochralski technique allows to crystallize up to 90\% of the loaded charge (already achieved for BGO~\cite{LTG_BGO}, CdWO$_4$~\cite{LTG_CWO1,LTG_CWO2} and ZnMoO$_4$~\cite{LTG_ZMO} crystals). This feature is crucial in case of costly material, like enriched isotopes or highly purified archaeological Pb.

\begin{figure}[t]
\includegraphics[width=0.45\textwidth]{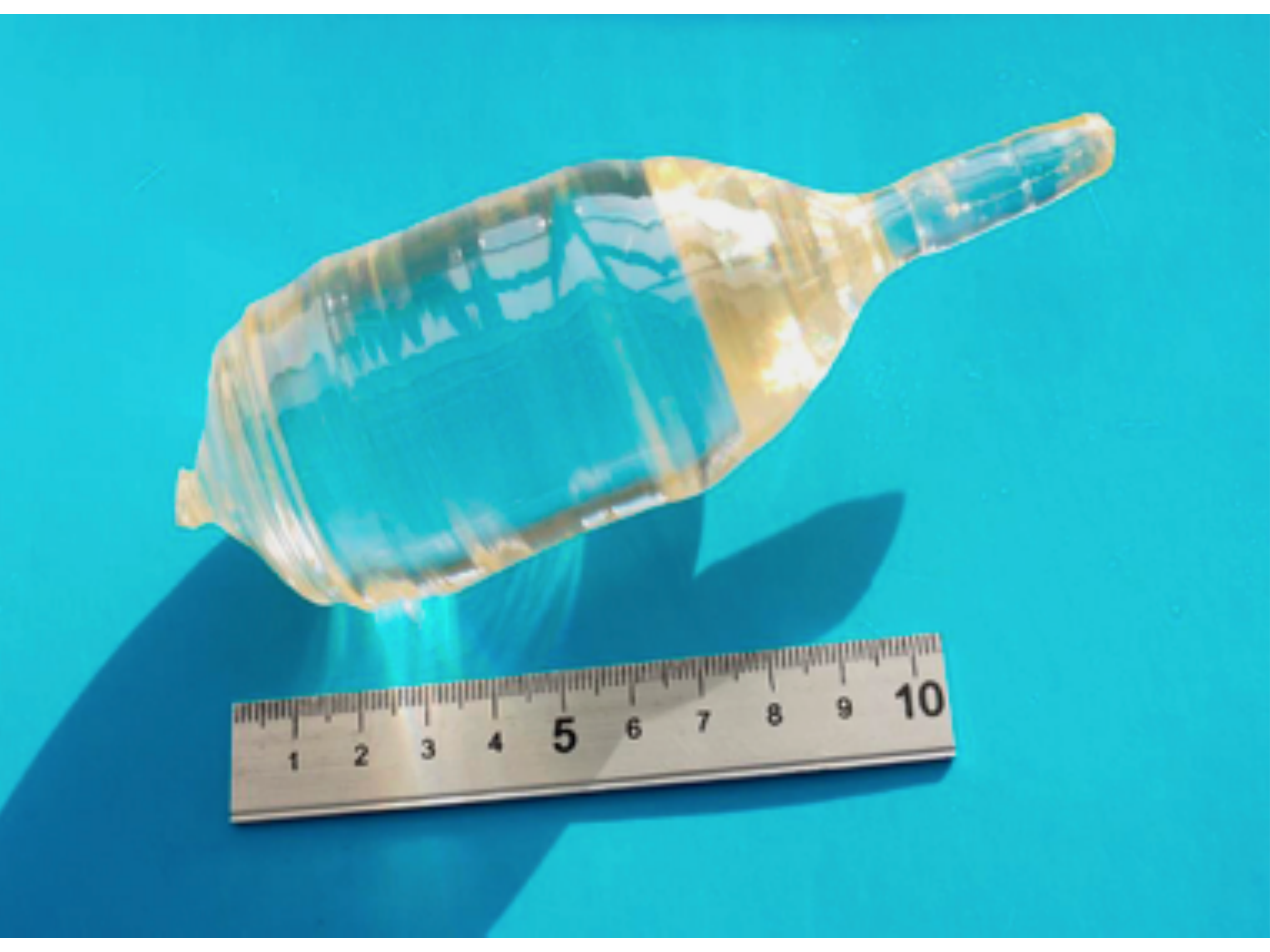}
\caption{Picture of a PbMoO$_4$ single crystalline boule with 911~g mass, the diameter and height of the boule are 45~mm and 100~mm, respectively. The crystal is grown by Low Thermal Gradient Czochralski technique using highly purified archaeological Pb.}
\label{fig:xtal}
\end{figure}

The crystal produced and characterized in this work was grown at the Nikolaev Institute for Inorganic Chemistry (Novosibirsk, Russia). The total mass of the loaded PbMoO$_4$ charge in the crucible was 1115~g, while the final single crystalline boule had a mass of 911~g (about 45$\times$100~mm). It indicates that total production yield was 82~\%, a little bit less than the expected value. In Fig.~\ref{fig:xtal} a photo of the obtained boule is shown.

The crystal was shaped into a cylinder with 44~mm diameter and 55~mm length, with a total mass of 570~g. The side surface was grinded, while the top and bottom ones were optically polished.

\section{Characterization as cryogenic detector}

The bolometric performance of the  PbMoO$_4$ crystal, its scintillation properties as well as its intrinsic radiopurity were investigated. The characterization of the compound was done in the low background cryogenic facility of CUPID R\&D in the underground laboratory of Gran Sasso (Italy). For details on the experimental set-up see~\cite{YVO}.
 
 \subsection{Detector operation}
The crystal was installed in a copper holder and fixed with PTFE clamps which acted as thermal insulator between the crystal and the thermal bath. The detector was installed in a dilution refrigerator and operated at a temperature of about 15~mK. The temperature variations induced by particle interactions were read-out by means of a Ge Neutron Transmutation Doped thermistor (Ge-NTD). Details on the experimental set-up can be found in~\cite{salaC,LiMoNe}.

PbMoO$_4$ proved to be also an efficient scintillator at low temperature~\cite{YRM}, for this reason a cryogenic light detector (LD) was faced to the absorber. This consisted of a thin wafer of Ge (170~$\mu$m thick) equipped with a Ge-NTD similar to the one used for the PbMoO$_4$ crystal. For more details on the operating condition of the LD refer to \cite{LD_perfomance}. The double read-out of heat and light allows for an efficient particle identification as already proven by many cryogenic scintillating calorimeters~\cite{YVO,ZS,LiEuBO}

The detector response was studied using different types of calibration sources: $^{228}$Th as $\beta$/$\gamma$ source, and  AmBe for neutrons. The internal radiopurity of the crystal was also investigated performing background runs, where no external calibration source was placed next to the experimental set-up.

\begin{figure}[t]
\centering
\includegraphics[width=0.45\textwidth]{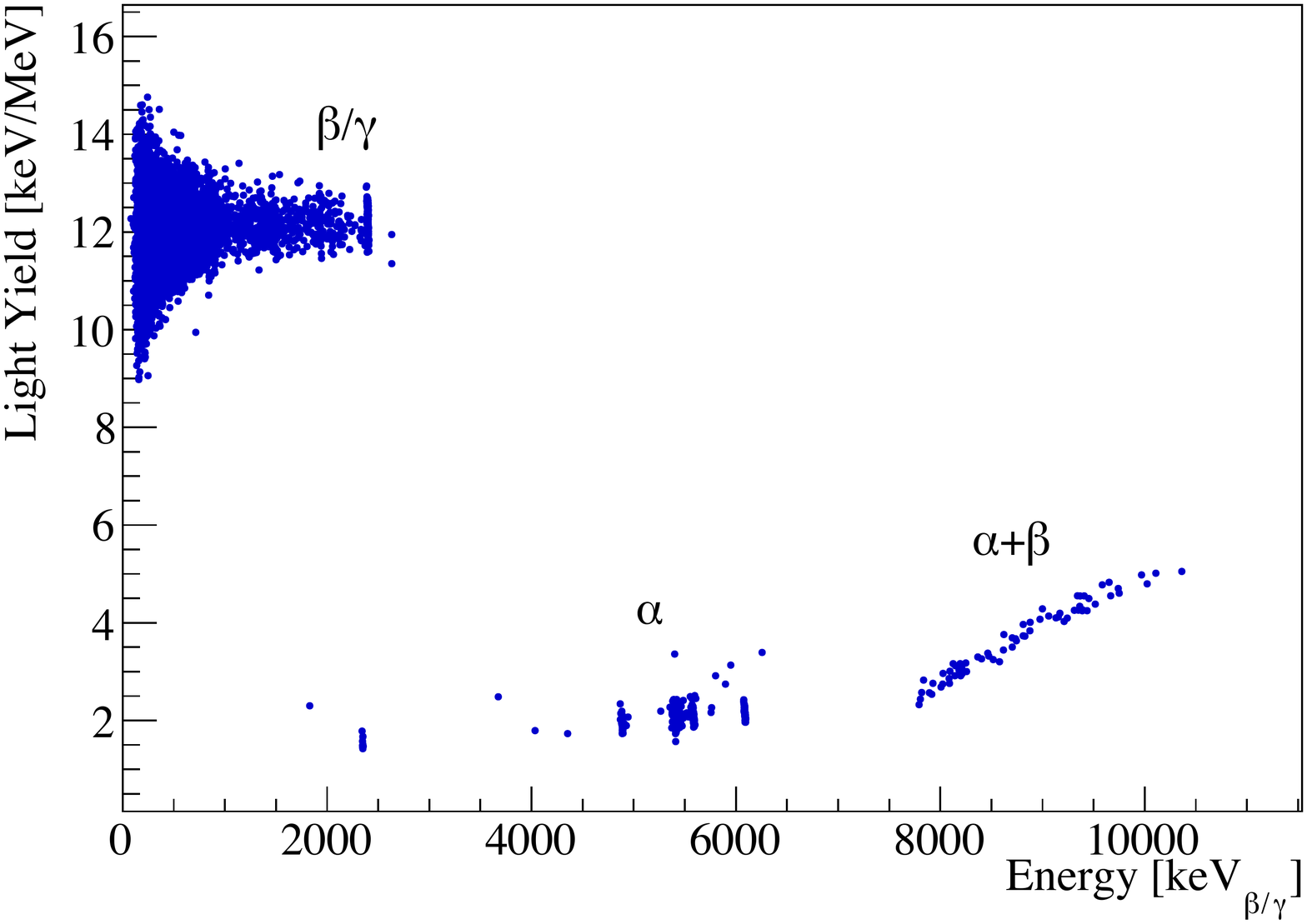}
\caption{Light yield vs. energy scatter plot of a 570~g PbMoO$_4$ scintillating crystal acquired over 47~h. The detector is exposed to a $^{228}$Th $\beta$/$\gamma$ calibration source. Three classes of events are highlighted, depending on the interacting particles.}
\label{fig:scatter_calib}
\end{figure}

A 47~h calibration run was performed deploying a $^{228}$Th source next to the experimental set-up. In Fig.~\ref{fig:scatter_calib} the acquired calibration energy scatter plot is shown, where the light yield\footnote{we define the light yield as the ratio between the measured light, in keV, and the nominal energy of an event, in MeV} (LY) is plotted as function of the energy deposited in the main absorber.

Three classes of events can be easily identified. The lowest energy class is ascribed to $\beta$/$\gamma$ interactions in the absorber. Then at higher energy there are $\alpha$ decays occurring in the crystal bulk, due to internal radioactive contaminations of $^{238}$U, $^{232}$Th and $^{147}$Sm. In some cases the decay of some daughter nuclei is so fast that cryogenic detectors do not have enough time resolution to disentangle the two events. This is the case of $\alpha+\beta$ events of Bi-Po decays, which produce pile-up events in the detector, the corresponding light signal is the sum of the light produced by the two independent events.

We computed the LY for $\beta/\gamma$ interactions to be 12.0$\pm$0.1~keV/MeV. While for $\alpha$ particle interactions the LY is 1.6$\pm$0.1~keV/MeV, for the lowest energy line produced by $^{147}$Sm $\alpha$-decay (at 2311~keV), and 2.1$\pm$0.1~keV/MeV for the most intense $\alpha$ line produced by $^{210}$Po decay (at 5407~keV). Given these values, we can compute the quenching factor for $\alpha$ particles, as described in~\cite{ZMO_small}, (QF$_{\alpha}$): 0.13$\pm$0.01 for  $^{147}$Sm and 0.17$\pm$0.01 for $^{210}$Po.

\begin{figure}[t]
\includegraphics[width=0.45\textwidth]{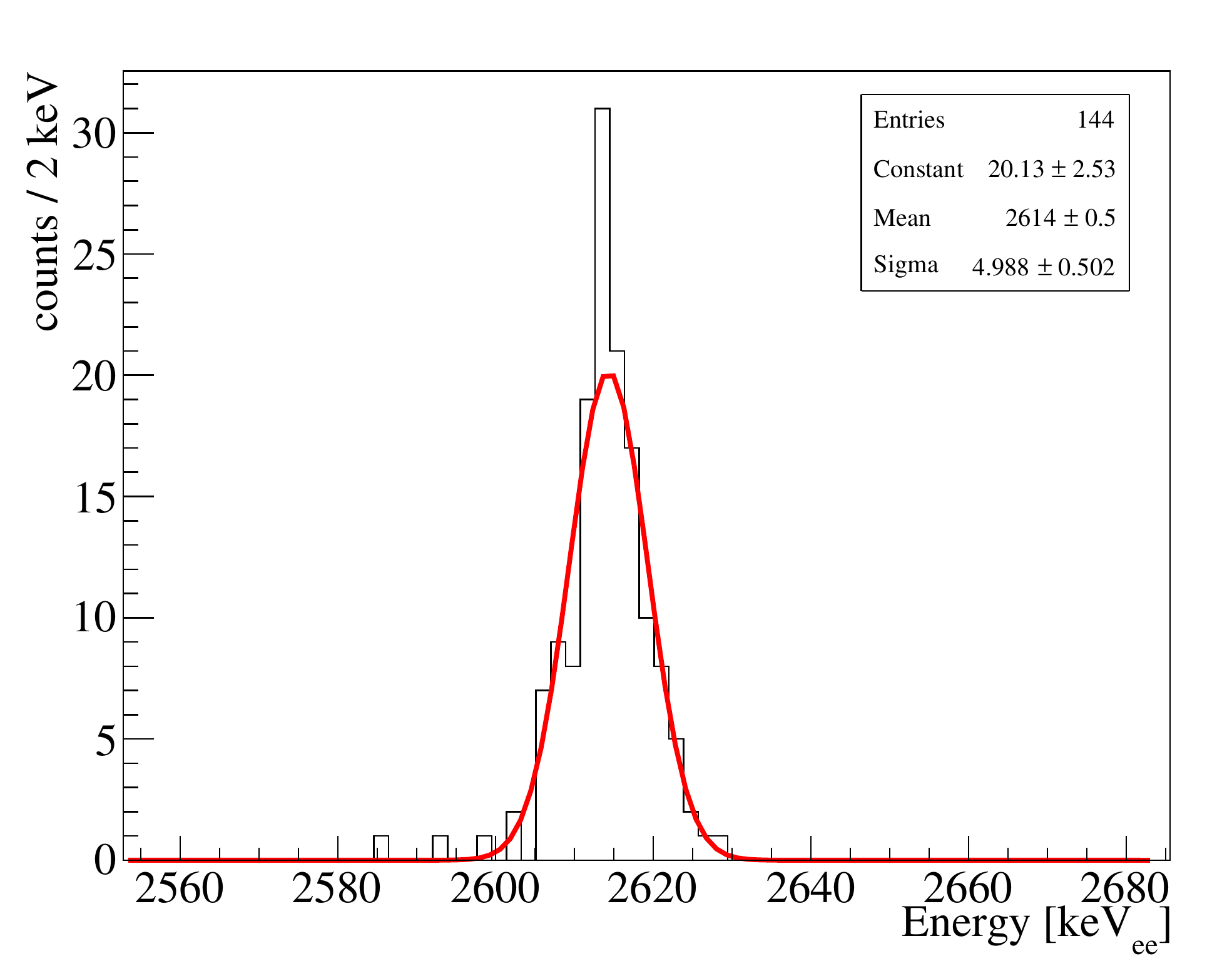}
\caption{Energy response of the detector to 2.6~MeV $\gamma$ interactions produced by the external $^{228}$Th calibration source.}
\label{fig:2615}
\end{figure}

In Fig.~\ref{fig:2615} the detector response at 2.6~MeV is shown. This line is fitted with a standard Gaussian function as typical for cryogenic detectors. The FWHM energy resolution for $\beta/\gamma$ interactions is 11.7$\pm$1.2~keV. This line is of great interest because it can be used as a proxy for the $^{100}$Mo $0\nu\beta\beta$ decay signature at 3.0~MeV. The achieved resolution is at the level of 0.2\%, an excellent value characteristic of cryogenic detectors. 

Finally in Tab.~\ref{Tab:Mo}, we show a direct comparison of the cryogenic performance of our detector with other highly performing Mo-based compounds.

\begin{table}
\caption{Comparison of the performance of the most promising Mo-based compounds for 0$\nu\beta\beta$ searches with cryogenic detectors. The light yield (LY) for $\beta/\gamma$ interactions, the quenching factor (QF) for $\alpha$ particles and the energy resolution around the region of interest are shown.}
\begin{center}
\begin{tabular}{|c|c|c|c|c|}
\hline
Crystal       & \begin{tabular}[c]{@{}c@{}}LY$_{\beta/\gamma}$\\ {[}keV/MeV{]}\end{tabular} & QF$_{\alpha}$ & \begin{tabular}[c]{@{}c@{}}FWHM$_{2615}$\\ {[}keV{]}\end{tabular} & ref.                                   \\ \hline
Li$_2$MoO$_4$ & 0.99& 0.22          & 6& \cite{LMO_exp}       \\ \hline
ZnMoO$_4$     & 1.54& 0.17          & 6.8& \cite{ZnMoO4_big}    \\ \hline
CaMoO$_4$     & -& -             & 6.3& \cite{CMO}            \\ \hline
CdMoO$_4$     & 2.55                                                                        & 0.16          & 13                                                                &\cite{CdMo} \\ \hline
PbMoO$_4$     & 12.0                                                                        & 0.17          & 11.7                                                              & This work                              \\ \hline
\end{tabular}
\label{Tab:Mo} 
\end{center}
\end{table}

\subsection{Crystal radiopurity}

\begin{figure}[t]
\includegraphics[width=0.45\textwidth]{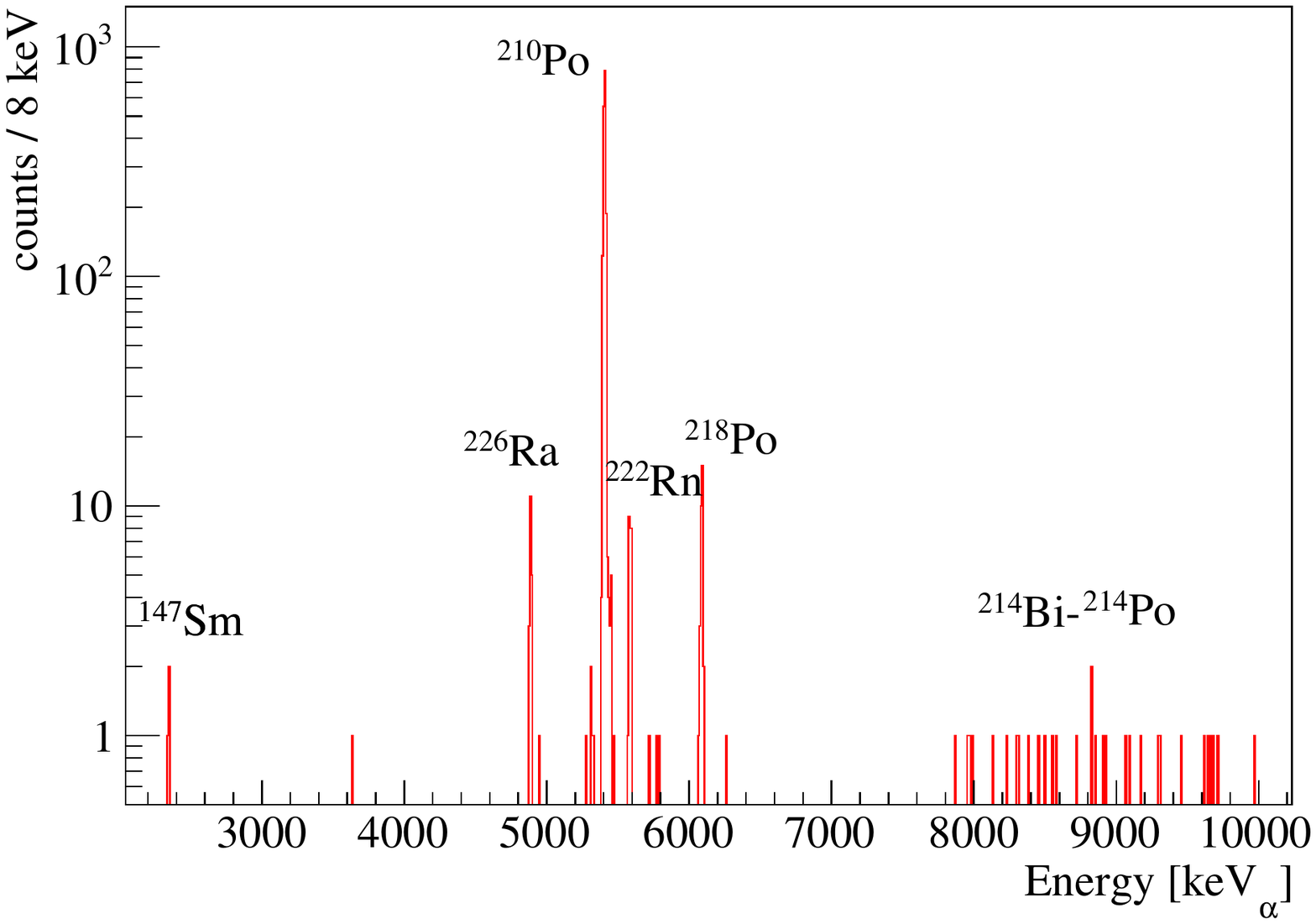}
\caption{Total background alpha energy spectrum of the 570~g PbMoO$_4$ crystal operated as cryogenic detector, acquired over 49~h.}
\label{fig:bkg}
\end{figure}

In order to have a better understanding of the internal radiopurity of the crystal a background run was performed. This was carried out in the same configuration as the calibration one, except for removing the external $\beta/\gamma$ source. In Fig.~\ref{fig:bkg}, the acquired background energy spectrum for 49~h of data taking is shown. Selecting the $\alpha$ band by a simple data selection cut on the LY, we can quantify the crystal radiopurity.

The radiopurity level of the crystal is still not competitive with other compounds employed for $0\nu\beta\beta$ searches~\cite{IHE}. Nevertheless the level achieved with the crystal studied in this work is the best nowadays for PbMoO$_4$ crystals, see Tab.~\ref{tab:cont}. The radionuclides which feature the highest concentration are $^{226}$Ra: 1.0$\pm$0.1~mBq/kg, $^{210}$Po: 53.0$\pm$0.1~mBq/kg and $^{147}$Sm: 123$\pm$62~$\mu$Bq/kg. No signature for $^{235/238}$U and $^{232}$Th parent nuclides could be identified. For these radionuclides only limits were set: $<$74~$\mu$Bq/kg. The Feldman-Cousins statistical approach was adopted for the computation of the limits. 

\begin{table}
\caption{Evaluated internal radioactive contaminations for the PbMoO$_4$ crystal. In the last column, the values for the first  PbMoO$_4$ crystal grown from archaeological Pb are shown~\cite{YRM} for comparison. Limits are computed at 90\% C.L.} 
\begin{center}
\begin{tabular}{lccc}
\hline\noalign{\smallskip}
Chain & Nuclide  & Activity & Activity \cite{YRM} \\ 
            & & [mBq/kg] & [mBq/kg]\\
\noalign{\smallskip}\hline\noalign{\smallskip}
$^{232}$Th & $^{232}$Th & $<$~0.074 & 10$\pm$7 \\
\noalign{\smallskip}\hline\noalign{\smallskip}
$^{235}$U & $^{235}$U & $<$~0.074 & 14$\pm$3\\
\noalign{\smallskip}\hline\noalign{\smallskip}
$^{238}$U & $^{238}$U & $<$~0.074 & 38$\pm$6\\
 & $^{226}$Ra & 1.0$\pm$0.1 & 2310$\pm$42\\
 & $^{210}$Po & 53.0$\pm$0.1& 266$\pm$16 \\
\noalign{\smallskip}\hline\noalign{\smallskip}
$^{147}$Sm & $^{147}$Sm & 0.12$\pm$0.06 & --\\
\noalign{\smallskip}\hline
\end{tabular}
\label{tab:cont} 
\end{center}
\end{table}

We can remark that the secular equilibrium among parent and daughter nuclides of the U/Th decay chains is broken, $^{226}$Ra is observed while $^{238}$U is not. This situation is rather typical for inorganic crystals, e.g. see~\cite{equilibrium}. These nuclides enter during the chemical processing of the raw material for the charge preparation, and, also later during crystal growth. The difference in their chemical properties (e.g. segregation effects) is the driving force that leads to different activities of nuclides of the same decay chain.

It can be observed from the background energy spectrum in the region of interest of $\alpha$ particle interaction, that the highest activity is ascribed to the decay of $^{210}$Po. $^{210}$Po has the highest concentration and could have been introduced by re-contamination of the initial charge before crystal growth.


Another point which deserves discussion is the presence of $^{147}$Sm, an unusual nuclide which was not observed in the raw materials used for the crystal growth process. This could be included in the crystal only during crystal growth, due to possible contamination of the crucible. This very same crucible was also used for the growth of other compounds.

\subsection{Particle discrimination on the heat channel}

The detector was also exposed to a neutron calibration source to produce high energy $\gamma$s needed to study the particle discrimination at higher energies. These are produced via (n,$\gamma$) reaction with the different materials in the experimental set-up (e.g. Cu, stainless steel, Au).
The ROI lies around 3~MeV, the region of interest for $^{100}$Mo $0\nu\beta\beta$ decay. Using these high energy $\gamma$s, a direct comparison between $\alpha$ and $\beta/\gamma$ events of the same energy is possible.

\begin{figure}[t]
\includegraphics[width=0.45\textwidth]{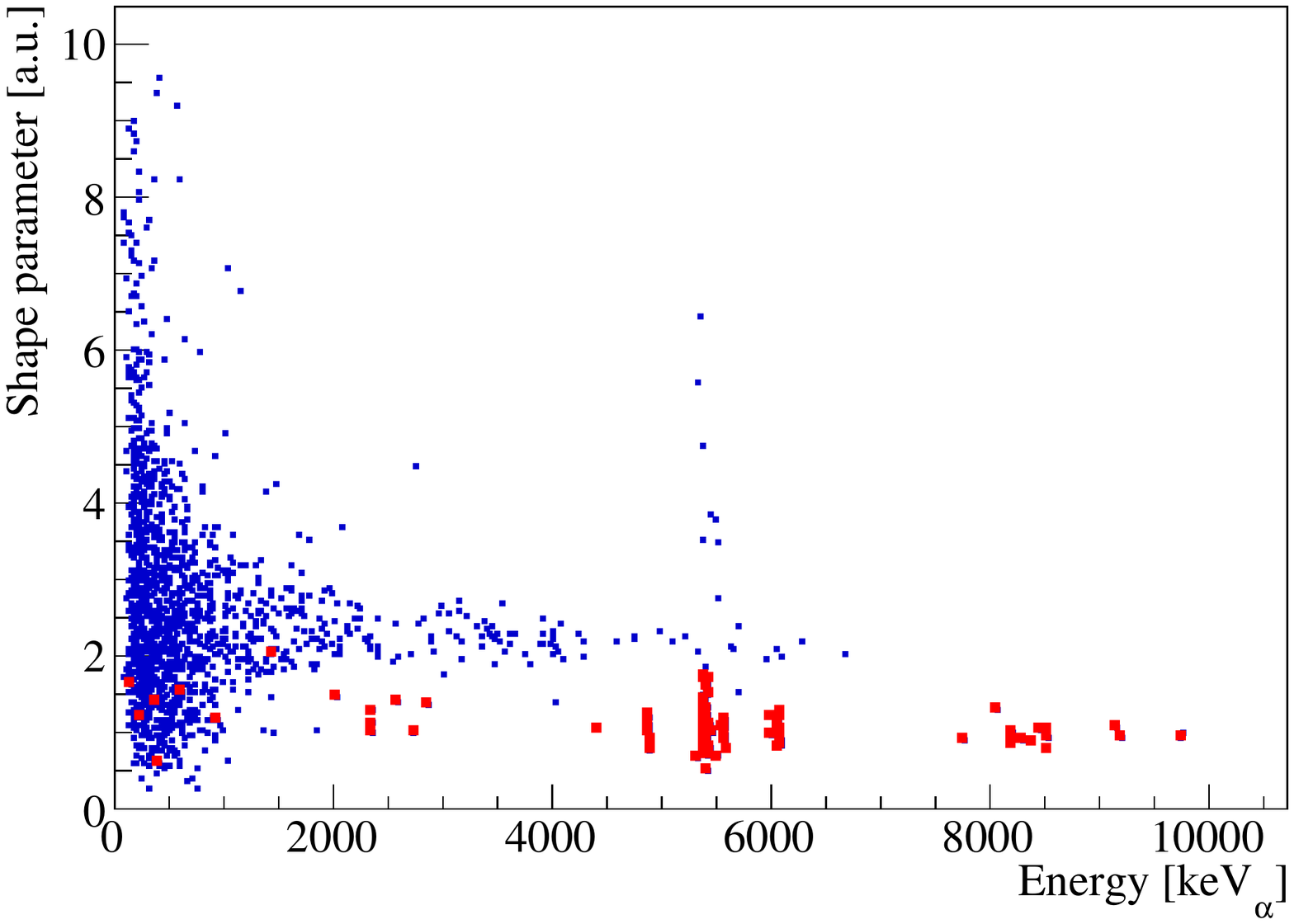}
\caption{Neutron calibration scatter plot for 15~h of data taking. The energy deposited in the PbMoO$_4$ is shown as a function of a shape parameter of the heat pulses. In red and in blue the $\alpha$ and the $\beta/\gamma$ events are highlighted, respectively.}
\label{fig:neutron}
\end{figure}

A neutron calibration scatter plot is shown in Fig.~\ref{fig:neutron}, the shape parameter (Test Value Right as described in~\cite{CUPID_analysis}) of the heat pulses is plotted as a function of the energy deposited in the PbMoO$_4$. The events highlighted in red are $\alpha$ interactions in the main absorber and in blue the $\beta/\gamma$ ones. If we look at the region between 2 and 5 MeV, a rather clear particle identification on the heat channel can be performed. This feature is characteristic of Mo-based compounds as already demonstrated for CaMoO$_4$~\cite{CMO_c}, Li$_2$MoO$_4$~\cite{LMO_exp} and ZnMoO$_4$~\cite{ZMO_small}. The driving mechanism for such behaviour relies on the slow scintillation decay time of Mo-based crystal (hundreds of $\mu$s) which has similar time constant as the time development of the heat signal of cryogenic detector. In fact, when luminescent centres are excited, these may de-excite via late phonon emission or scintillation light emission. The secondary phonon emission, together with the primary phonon production induced by the particle interaction, create a distortion of the thermal signal of the main absorber. Depending on the type of particle interaction (i.e. amount of produced light), we have different shape for the heat signal of the cryogenic absorber.  Details can be found in~\cite{LucaG}.

A dedicated measurement with higher statistical significance would allow an accurate and precise evaluation of the particle discrimination potential. The purpose of this work was to highlight only the most interesting feature of this compound.

This unique feature, namely particle identification without the need to operate a LD, enables a strong simplification of experimental set-ups where only one type of detector can be operated. This is mostly relevant in view of large array of scintillating bolometers for future 0$\nu\beta\beta$ decay investigations.
 
 \section{Conclusions}
We operated a PbMoO$_4$ as scintillating cryogenic calorimeter in the deep underground laboratory of Gran Sasso (Italy). The detector was produced from archaeological Pb, a unique material for the realization of low background detectors. The concentration of impurities, chemical and radioactive, inside the crystal were investigated. Furthermore, the properties as cryogenic detector were also studied, namely the light emission, its response to different types of radiation and the energy resolution.

The crystal shows an extremely good purity level, no major impurities were measured (at 10$^{-6}$~g/g scale). The only relevant contamination is W, which does not affect the detector performance. Thanks to the low background environment in which the detector was operated, we were able to evaluate the concentration of radionuclides inside the crystal. The only measurable contaminants are $^{226}$Ra and $^{210}$Pb, from the $^{238}$U decay chain, and $^{147}$Sm. Most probably these are included in the crystal during the growth process, where no precautions were taken to prevent re-contamination of the highly-pure raw materials. In fact, according to~\cite{newPbRo}, archaeological Pb features excellent radiopurity, as well as MoO$_3$~\cite{MoO3}. If we consider only the radiopurity level of the raw materials, this crystal could be among the best candidates for the realization of a 0$\nu\beta\beta$ cryogenic experiment. 
Nevertheless, a thorough control of the different production processes (e.g. powder synthesis, crystal growth) is mandatory for the realization of crystal with a radiopurity competitive with other 0$\nu\beta\beta$ compounds.

In this work we also studied the light emission at mK temperature. The crystal features an intense light emission for $\beta/\gamma$ particles of 12~keV per 1~MeV of energy deposit, the highest among Mo-based compounds. This characteristic is of undoubtable importance for performing an efficient particle interaction discrimination. We also estimated the quenching factor for $\alpha$ interactions, and it concurs with other value reported in literature for Mo-based compounds.

The cryogenic properties of the detector were excellent. We computed the detector FWHM energy resolution at 2.6~MeV to be 11.7~keV. This value is competitive with other compounds used as cryogenic detector for $\beta\beta$ decay searches. Possibly a strict control of the production processes may lead to a higher crystal purity, which will positively affect the cryogenic performance of the detector (lower metal impurity concentration results in larger signal amplitude).

Another important feature investigated, while operating this compound, is the particle identification using pulse shape analysis on the heat channel. This is a unique feature of Mo-based compounds, which allows for a simplification of the experimental set-up, where no LD needs to be operated. Further dedicated studies are needed for quantifying the discrimination potential, as well as for studying the dependance of the discrimination on the operating temperature of the detector.

\begin{acknowledgements}
Thanks are due to the LNGS mechanical workshop and in particular to E.~Tatananni, A.~Rotilio, A.~Corsi, and B.~Romualdi for continuous and constructive help in the overall set-up construction. Finally, we are especially grateful to M.~Perego and M.~Guetti for their valuable help.
\end{acknowledgements}

\end{document}